# All-condition pulse detection using a magnetic sensor

Haoxuan Shen[1,2], Xiaoyi Gu[1] and Yihong Wu[1,2,a]

*[1]Department of Electrical and Computer Engineering, National University of Singapore, Singapore 117583, Singapore*

*[2]National University of Singapore (Chong Qing) Research Institute, Chongqing Liang Jiang New Area, Chongqing 401123, China*

A plethora of wearable devices have been developed or commercialized for continuous non-invasive monitoring of physiological signals that are crucial for preventive care and management of chronic conditions. However, most of these devices are either sensitive to skin conditions or its interface with the skin due to the requirement that the external stimuli such as light or electrical excitation must penetrate the skin to detect the pulse. This often results in large motion artefacts and unsuitability for certain skin conditions. Here, we demonstrate a simple fingertip-type device which can detect clear pulse signals under all conditions, including fingers covered by opaque substances such as a plaster or nail polish, or fingers immersed in liquid. The device has a very simple structure, consisting of only a pair of magnets and a magnetic sensor. We show through both experiments and simulations that the detected pulsation signals correspond directly to the magnet vibrations caused by blood circulation, and therefore, in addition to heartrate detection, the proposed device can also be potentially used for blood pressure measurement.

Continuous non-invasive monitoring of physiological signals such as heart rate (HR), blood pressure (BP) and blood oxygen level are crucial for preventive care and management of chronic conditions[1,2]. A plethora of devices have been developed for this purpose, notably photoplethysmography (PPG) which can detect the blood volume changes in the microvascular bed of tissue and arteries through measuring the intensity variation of either transmitted or reflected light from the tissue[3]. From the detected optical signals, we can extract physiological signals such as heart rate and blood oxygen level[4]. Due to its high cost-performance and ease of use, the PPG has become the *de facto* standard non-invasive vital sign monitoring device in both clinical and non-clinical settings. Despite the high penetration rate, the PPG has several drawbacks such as motion artifacts, skin tone effect, and unsuitability for circumstances where the skin is covered by opaque materials (e.g., plaster or nail polish). Therefore, there is always a need to develop non-optical detectors which can replace or supplement the PPG.

One of the promising techniques that can potentially overcome some of the difficulties facing by the PPG is the magnetoplethysmography (MPG)[5,6]. The early work on MPG was motivated by the fact that oxygenated haemoglobin (HbO$_2$) and de-oxygenated haemoglobin (Hb) exhibit different magnetic properties; the former is diamagnetic whereas the latter is of paramagnetic nature[7]. Phua et al. have demonstrated a wrist-type MPG device consisting of a permanent magnet and a Hall sensor which can clearly detect the pulsatile signals in both in-vitro and in-vivo configurations[5,6]. The authors attributed the MPG signals to modulated magnetic signature of blood (MMSB)[6,8], based on the hypothesis that the magnetic field of permanent magnet affects the blood flow dynamics, which in turn perturbs the magnetic field distribution in the surrounding area. Since then several research groups have demonstrated MPG prototype devices using either the giant magnetoresistance (GMR) or Hall sensor which successfully detected the pulsatile signals, demonstrating good reproducibility of the detection method[9-16]. Considering the difference in magnetic responses of $HbO_2$ and Hb, the initially proposed MMSB mechanism seems plausible. However, given the small difference (on the order of several ppm) in magnetic susceptibility between $HbO_2$ and Hb[7], it remains controversial whether the MPG signal originates entirely or in part from the MMSB mechanism. Simulations by Sinatra revealed that the disturbance to magnetic field caused by the magnetic properties of blood is on the order of $10^{-5}$, much smaller than the earth's magnetic field[17], although it shows close correlation with the blood flow velocity. The in-vitro experiments performed by Zhang et al. suggest that the MPG signal is mainly derived from the mechanical disturbance of the sensor detection axis caused by blood flood, but quantitative discussion is lacking[18]. Similar results were also reported by Li et al., though it is not clear if the results were obtained from in-vivo or in-vitro measurements[19]. Without a proper understanding of the detection mechanism, it would be difficult to further develop the MPG for practical applications.

Here, we propose a fingertip-type magnetic device which allows to measure the magnetic and vibrational signals simultaneously using a magnetic sensor and a laser Doppler vibrometer. Unlike all previous experiments which place a single magnet on the radial artery, we place two magnets on the opposite sides of a fingertip where the mutual attraction between the magnets significantly improves the stability of both magnetic and vibrational signals. We found that the magnetic and vibrational signals are closely correlated with each other, and the average deviation between the two signals is less than 5%. Since the magnetic signal is diminishingly small when the magnets are not in contact with the skin, we can safely conclude that the detected signal is dominated by vibration of the magnet, and the contribution from the MMSB mechanism is negligible. Both analytical and numerical simulation results are in good agreement with the experimental observations. We show that such kind of device can detect clear pulsation signals under all finger conditions – a task which has proven to be rather challenging for existing pulse detectors. Since the measured signal is not originated from the MMSB mechanism, hereafter we call it magnetically detected vibration sensor (MDVS) instead of MPG.

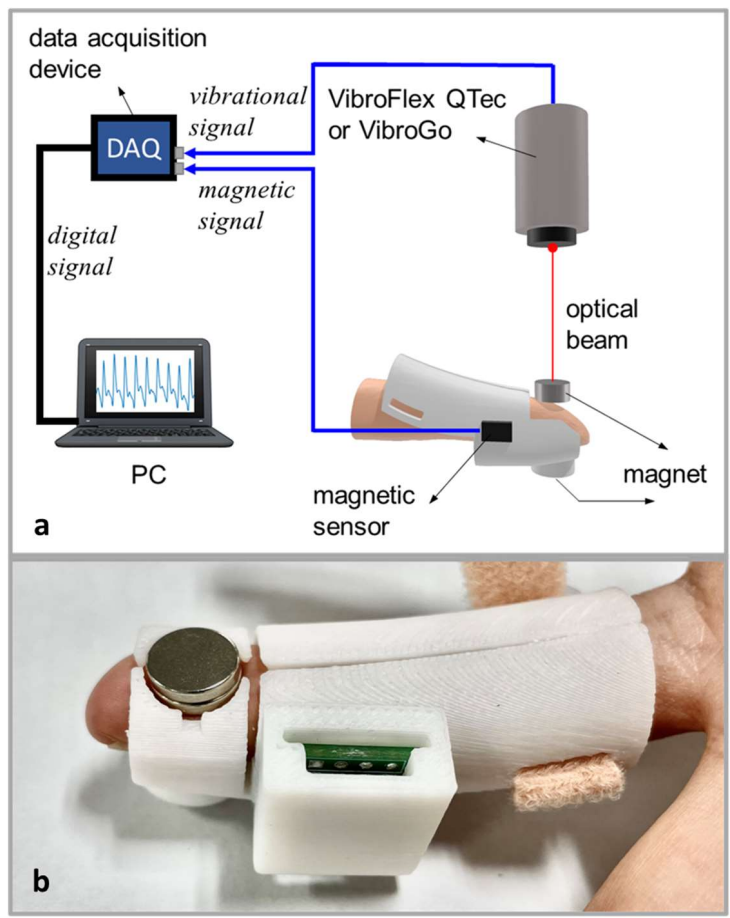

**FIG.1.** Schematic of the fingertip-type MDVS device and measurement setup. (a) Measurement set up. (b) Photograph of assembled fingertip-type MDVS.

**Results**

The MDVS device used in this work consists of two cylindrical magnets attached to a fingertip, one of which is firmly attached to a mechanical fixture and the other is movable in response to blood flow-induced skin movement (Fig.1). A magnetic sensor is placed near the movable magnet (approximately 1.5 cm), which detects changes in the magnetic field of the magnet. Depending on the magnet chosen and positioning of the sensor, various types of magnetic sensors can be used, including the recently developed spin-orbit torque enabled magnetic sensor[20-22]. To facilitate mounting of the sensor, the data presented in this paper are collected using a commercial tunnel magnetoresistance sensor with a dynamic range around 100 Oe and sensitivity of 1.28mV/Oe. The prototype is designed such that the movable magnet can be directly probed by a laser Doppler meter (Polytec VibroGo). The VibroGo signal acquisition unit has two input channels which can be used to measure the vibration and magnetic signals simultaneously. This greatly facilitates the comparison of two signals as the time difference caused by measurement electronics is presumably negligible. The experiments were performed on multiple fingers of healthy subjects of different gender and age. Good reproducibility has been obtained in all measurements conducted.

**Synchronized detection of magnetic and vibration signals and statistical analysis**. Figure 2 shows the typical magnetic and vibrational signals obtained simultaneously from two healthy subjects. Figure2a is the pulsatile signal of subject A (male at late 50's) measured by a magnetic sensor for a duration of 16 seconds. The aspiratory baseline shift has been removed and the data have been smoothed with a bandpass filter with a bandwidth of 0.8~10 Hz. Comparison with commercial PPG sensor confirms that the periodicity of the signal is corresponding to the heartbeat rate of the subject. Figure 2b shows the corresponding average single pulse signal with the mean subtracted out and normalized by the standard deviation. The overall shape resembles well the PPG waveform with clear systolic peak, dicrotic notch and diastolic peak[3]. A high degree of similarity with PPG is also seen in the acceleration pulse waveform (or second derivative), as shown in Fig.2c, which is useful for evaluating cardiovascular conditions[23]. Figure 2d-2f show the same set of data for subject B (female at early 20's). The basic characteristics are the same as those obtained from Subject A, though we do see some subtle differences in the shape of the pulse and acceleration waveform. These results demonstrate clearly that the magnetic signals are reproducible and of good quality. We now turn to the vibrational signals acquired simultaneously by VibroGo from subject B. The pulsatile signals showing in Fig.2g and 1h are the time-dependent displacement signals, whereas Fig.2i is the acceleration signal measured directly by the Doppler meter. Apparently, there is a close correlation between the magnetic and vibrational signals in terms of both periodicity and waveform including the characteristic peaks. The subtle difference in systolic peak position is presumably due to the different time response of the two measurement channels. The correlation coefficient is in the range of 0.9527 - 0.9924, which is even higher for the averaged single pulse.

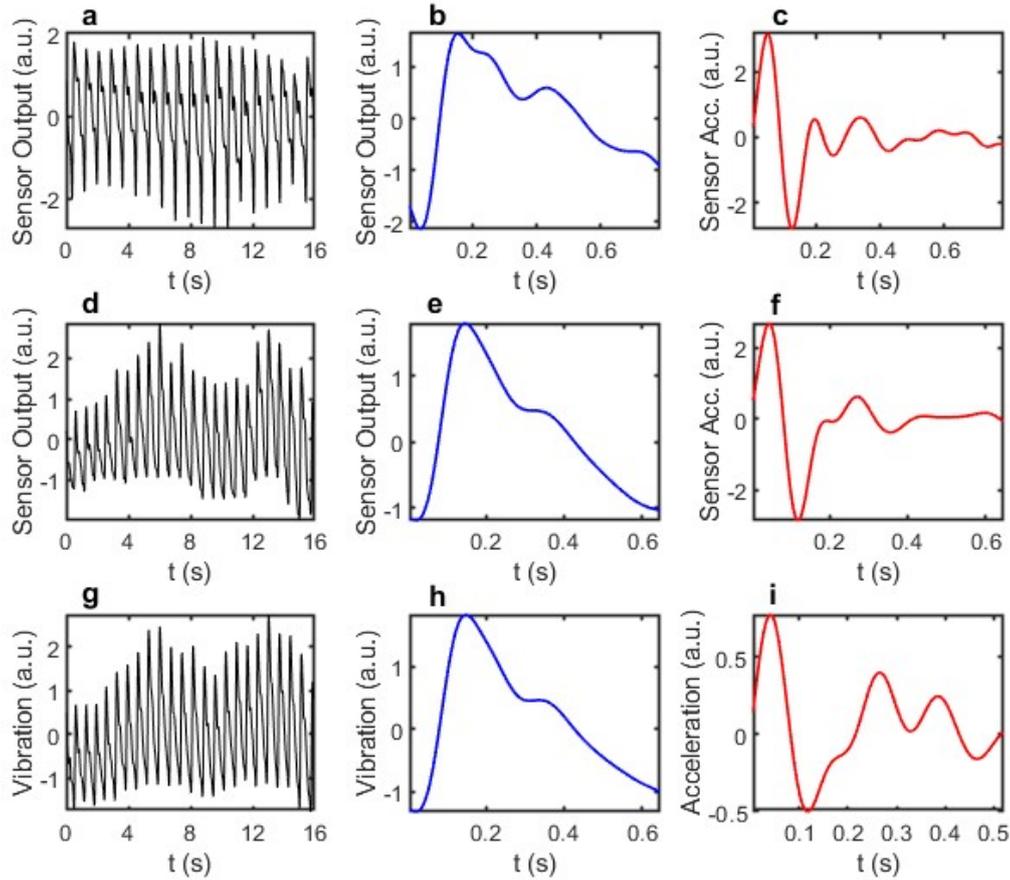

**FIG.2.** Magnetic and vibrational signals obtained from two healthy subjects. (a) - (c) are signals obtained from subject A: (a), magnetic sensor output signal for a duration of 16s, (b), average single pulse signal from (a), and (c), second derivative of (b). (d) - (f) are the corresponding magnetic signals measured from subject B. For comparison, (g) - (i) shows the vibrational signals measured simultaneously with (d) - (f): (g) is the time-dependent displacement signal, (h) is the average single pulse signal and (i) is the average acceleration detected directly by the vibrometer.

**Pulse monitoring under multiple conditions.** To show the 'all-condition' pulse monitoring capability of the device, we have conducted the measurements for fingers under different conditions: (a) wrapped by a black tape, (b) wrapped by a plaster, and (c) immersed in water, and the results are shown in Fig. 3a-3c, respectively. As shown in the left panel of Fig. 3a and 3b, the contact area of the skin and magnet is completely covered by a black tape or a plaster. The former completely blocks the light whereas the latter is semi-transparent. Should a PPG sensor be used for the detection, one would not expect any measurable signal in (a) or a clear signal in (b). But, as shown in the right panel of Fig. 3a and 3b, clear signals are obtained in both cases, indicating that the black tape or plaster has very little effect to the MDVS detector. To demonstrate that the MDVS is immune to body fluid too, we conducted the measurement by immersing the fingertip and magnets in water, while leaving the magnetic sensor in air (since it is not made waterproof). As shown in the right panel of Fig.3c, again we can obtain very clear pulse signals. Although the test was done using clear water, we expect that similar results can be obtained when there is body fluid such as sweat and blood on the finger. Repeated experiment shows that the signals detected under all the three conditions are stable and reproducible, with clear systolic peak, dicrotic notch, diastolic peak, and equal peak-to-peak span. There is very little difference compared to the results obtained from bare hand in air. These results clearly demonstrate the advantages of MDVS over existing pulse detectors such as PPG, in coping with challenging circumstances.

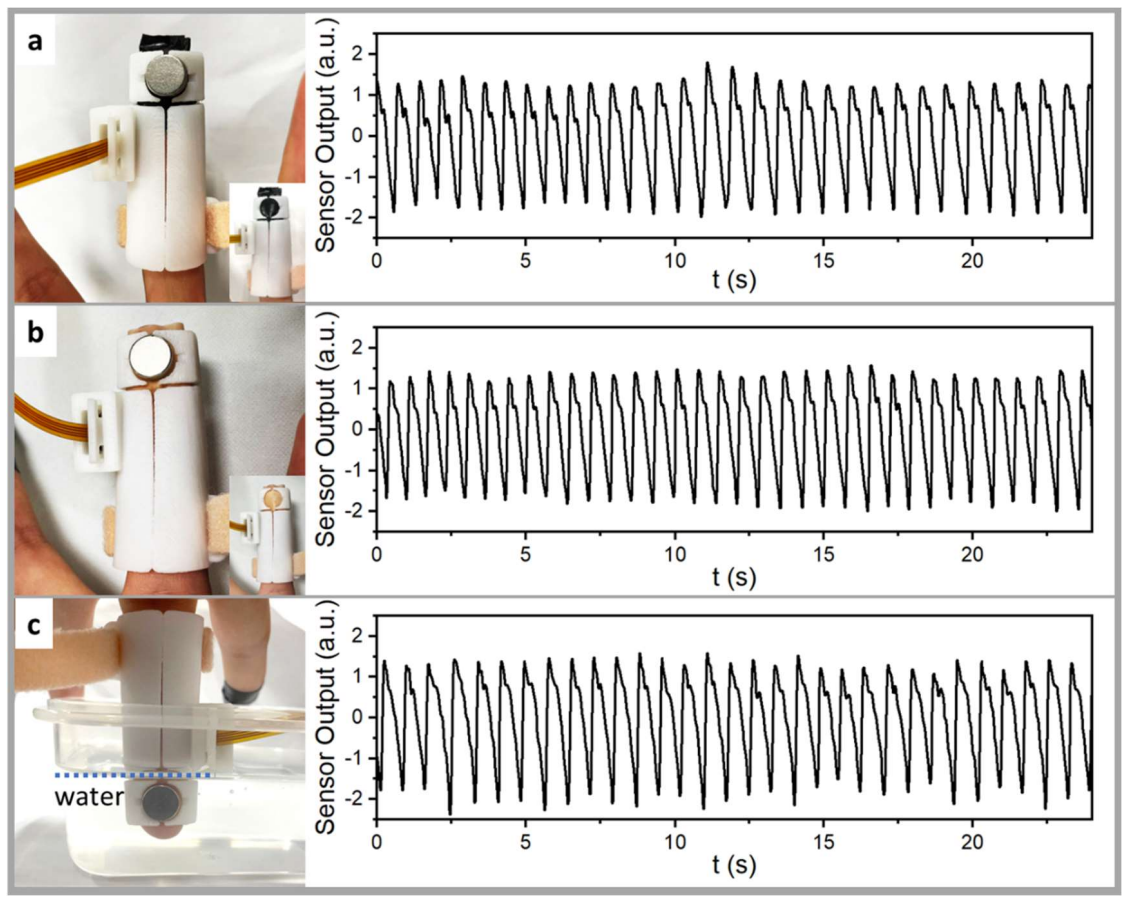

**FIG.3.** Results of pulse measurements with the fingertips (a) wrapped by a black tape, (b) wrapped by a plaster, and (c) immersed in water. Left panel: photography of the MDVS and fingertip during measurement; right panel: detected signals.

**Bland-Altman plot analysis of magnetic and vibration signal.** To have a more quantitative comparison, we compare normalized magnetic and vibrational data and perform Bland-Altman analysis in Fig.4. The Bland-Altman plot is often used to analyze the matching between two methods of measurements for the same variable. It plots the difference against the mean of two sets of measurement values, highlighting the degree of agreement between the two measurement methods. The recommended limit of agreement is that 95% of the data points should lie within ±1.96 times of standard deviation (SD) of the mean difference[24,25]. Figure 4a and 4b show the average single pulse of magnetic (blue solid-line) and vibrational signal (red solid-line) measured simultaneously from subject B. The vibrational signals are obtained using two different models of vibrometers: VibroFlex QTec for Fig.4a and VibroGo for Fig.4b. Other than a slight difference in surface roughness requirements, there are no key technical differences between the two models of vibrometers. The main purpose of respectively. (c) and (d), Bland-Altman plots based on repeated measurements using VibroFlex QTec (c) and VibroGo (d).

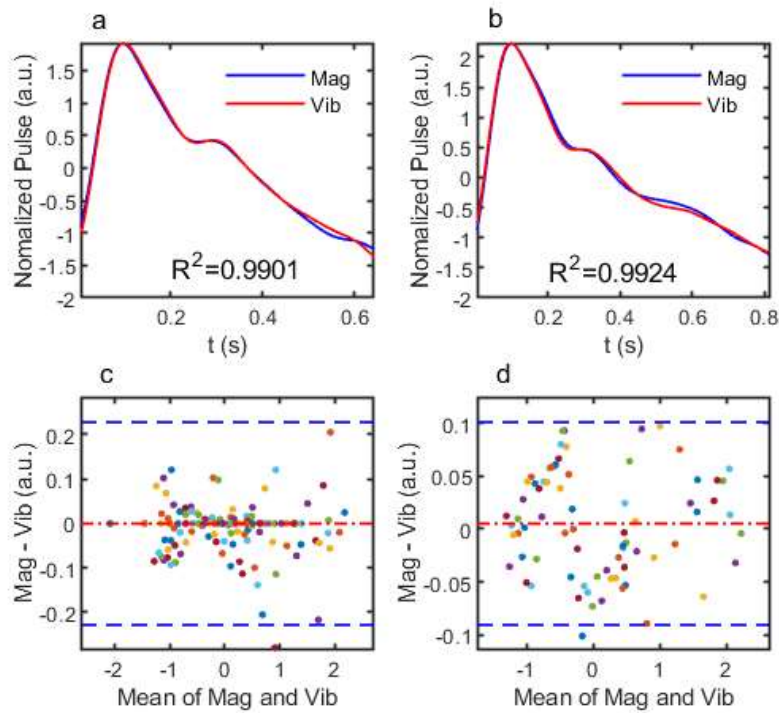

**FIG.4.** (a) and (b) comparison of single pulse signal magnetic and vibrational signals. The vibrational signals are measured using VibroFlex QTec (a) and VibroGo (b),

using two different models of vibrometer is to check the correlation between magnetic and vibrational signals. The correlation coefficient between magnetic and vibrational signals in Fig.4a is 0.99, indicating the same origin of the two signals. To quantify the difference statistically, we add more datasets to produce the Bland-Altman plot in Fig.4c. The two dashed-lines denote ±1.96 SD. Except for a few outliers, the majority of the data points fall within the difference range of - 0.23 to 0.22, with a mean bias of - $2.5 \times 10^{-4}$. Similar results are obtained for signals measured by the VibroGo, with a correlation coefficient of 0.99, and difference between -0.09 and 0.1 and mean bias of $5.1 \times 10^{-3}$ in the Bland-Altman plot (Fig.4d). The maximum deviation between the two measurement methods is 7.12% and 2.99%, respectively, for vibration measurements using VibroFlex QTec and VibroGo. In addition to the use of different types of vibrometer, we have also repeated the measurements using different types of magnetic sensors. These results suggest that the close correlation between magnetic and vibrational signal is a generic phenomenon which does not depend on the vibrometer or magnetic sensor used to measure the signals.

**Analytical model of MDVS sensor.** In what follows, we examine more quantitatively the correlation between magnetic and vibrational signals using both analytical models and simulation. As discussed in the introduction, the MPG signal was initially interpreted as magnetic signature induced by blood flow. Later studies suggest that change of magnetic sensor's detection axis caused by blood flow may also play a role, but quantitative understanding is lacking. In the present setup, the magnets are attached to a fingertip which consists of nail, skin, muscle, bones, arterial/vein vessels, and capillaries[26]. When the blood flows through the vessels, it generates pressure wave on the vessel walls[27], which can propagate through the thick tissue and cause vibration of the skin. Therefore, without losing generality, the output signal of the magnetic sensor may be expressed as

$$\Delta V_{mag} = S\hat{s} \cdot \left(\frac{\partial \vec{H}}{\partial (\vec{r}_s - \vec{r}_m)} \Delta(\vec{r}_s - \vec{r}_m) + \frac{\partial \vec{H}}{\partial \vec{m}} \Delta \vec{m} + \frac{\partial \vec{H}}{\partial \chi} \Delta \chi\right) \qquad (1)$$

where $S$ is the sensitivity of the sensor and $\hat{s}$ is its sensing axis direction, $\vec{H}$ is the magnetic field at the sensor location, $\vec{r}_m$ ($\vec{r}_s$) is the position of the magnet (sensor), $\vec{m}$ is the magnetic moment of the magnet, $\chi$ is the magnetic susceptibility of the blood, and $\Delta(\vec{r}_s - \vec{r}_m)$, $\Delta \vec{m}$, and $\Delta \chi$ are the corresponding changes induced by the blood flow. The first two terms of Eq. (1) are corresponding to signals induce by parallel displacement and rotation of the magnet, respectively, whereas the last term is due to change in magnetic susceptibility of the blood. Before performing numerical simulations to calculate the overall signal, it is instructive to examine the first two terms analytically using the magnetic dipole approximation, which is valid when the distance between the magnet and sensor $|\vec{r}_s - \vec{r}_m|$ is much larger than the size of the magnet.

**Signal due to magnet displacement.** We first consider the effect of parallel displacement of the magnet, i.e., $\Delta \vec{m} = 0$. For simplicity, we assume that the sensor is stationary, and only one of the magnets is moving, which describes well the actual measurement setup. In this case, we may denote the magnet position as $\vec{r} = (x, y, z)$ in a Cartesian coordinate system with its origin at the sensor position. Then, the output signal may be written as $\Delta V_{mag}^D = S\hat{s} \cdot \frac{\partial \vec{H}}{\partial \vec{r}} \Delta \vec{r}$, where $\vec{H} = \frac{1}{4\pi}\left(\frac{3\vec{r}(\vec{m} \cdot \vec{r})}{|\vec{r}|^5} - \frac{\vec{m}}{|\vec{r}|^3}\right)$ and $\Delta \vec{r} = (\Delta x, \Delta y, \Delta z)$. When the detection axis of the sensor and diploe moment of the magnet

are aligned along x- and z-axis, respectively, *i.e.*, $\hat{s} = (1,0,0)$, $\hat{m} = (0,0,1)$, $\Delta V_{mag}^{D}$ is calculated as

$$\Delta V_{mag}^{D} = \frac{3mS}{4\pi} \frac{1}{r^7}[(x^3 + xy^2 - 4xz^2)\Delta z + (z^3 + zy^2 - 4zx^2)\Delta x - 5xyz\Delta y], \quad (2)$$

where $r = (x^2 + y^2 + z^2)^{1/2}$. When $x \gg z, y$, which represents the actual experimental setup, $\Delta V_{mag}^{D}$ is approximately given by $\Delta V_{mag}^{D} \approx (3mS/4\pi)(\Delta z/x^4)$. On the other hand, when the laser beam is aligned in z-direction, $\Delta z$ is nothing else but the displacement measured by the vibrometer. Therefore, the first term of Eq. (1) is directly proportional to the vibration signal. The difference, if any, is mainly caused by the transverse movement of the magnet, *i.e.*, $\Delta x$ and $\Delta y$, with respect to the laser beam direction, which can be minimized by optimizing the sensor position with respect to the magnet, *i.e.*, to make $x \gg z, y$.

**Signal due to magnet rotation.** Next, we consider the contribution due to rotation of the dipole which is initially aligned in z-direction, *i.e.*, $\vec{m} = (0,0,1)m$. Without losing generality, we assume that the yaw, pitch and roll angles are $\alpha$, $\beta$, and $\gamma$, respectively, and the change of $\vec{m}$ due to rotation is given by $\Delta \vec{m} = R(\alpha, \beta, \gamma)\vec{m} - \vec{m}$, where $R(\alpha, \beta, \gamma)$ is the rotation matrix. The signal induced by the dipole rotation can be calculated directly from Eq. (1) which reads

$$\Delta V_{mag}^{R} = \frac{3S}{4\pi} \frac{1}{r^5}\left[\left(x^2 - \frac{r^2}{3}\right)\Delta m_x + xy\Delta m_y + xz\Delta m_z\right], \quad (3)$$

where $\Delta m_x$, $\Delta m_y$, and $\Delta m_z$ are the changes of magnetic moment in the three coordinate axis directions. When $\alpha, \beta,$ and $\gamma$ are very small and $x$ is much larger than $z$ and $y$, $\Delta V_{mag}^{R}$ is approximately given by $\Delta V_{mag}^{R} = (Sm/2\pi)(\beta/x^3)$. The result shows that the magnetic signal induced by the rotation is directly proportional to the pitch angle $\beta$, *i.e.*, the angle of rotation around y-axis. Although the dipole approximation is valid in calculating the magnetic signal, to estimate the displacement signal measured by the vibrometer, we have to consider the finite size of the magnet. Here, we assume that the laser beam is initially focused at $\vec{L} = (L_x, L_y, L_z)$ on the top surface of the magnet with respect to the rotational center and its direction is parallel to z-axis. In this case, the displacement in z-direction is given by $\Delta L_z = \hat{k} \cdot (R(\alpha, \beta, \gamma)\vec{L} - \vec{L})$ and when $\alpha, \beta,$ and $\gamma$ are small, it reduces to $\Delta L_z \approx \gamma L_y - \beta L_x$. Therefore, the displacement signal contains two components which are proportional to the pitch and roll angles, respectively. In other words, depending on the rotation directions, the displacement measured by the vibrometer may not be exactly the same as that of the output signal from the magnetic sensors. The ratio between $\Delta V_{mag}^D$ and $\Delta V_{mag}^R$ is on the order of $(\Delta z/x)/\beta$. Based on the displacement measured by the vibrometer, $\Delta z/x$ is on the order of $10^{-3} - 10^{-2}$, which corresponds to a rotation angle of $0.06^o - 0.6^o$, which means that $\Delta V_{mag}^R$ can be comparable to or even larger than $\Delta V_{mag}^D$, depending on the particular measurement setup. This may explain why the agreement between measured magnetic and vibration signals is generally better in the systolic than the diastolic half-cycle of the heartbeat (Fig.2 and Fig.4), as the sudden decrease of blood pressure after the systolic peak may result in not just parallel displacement but also rotation of the magnet.

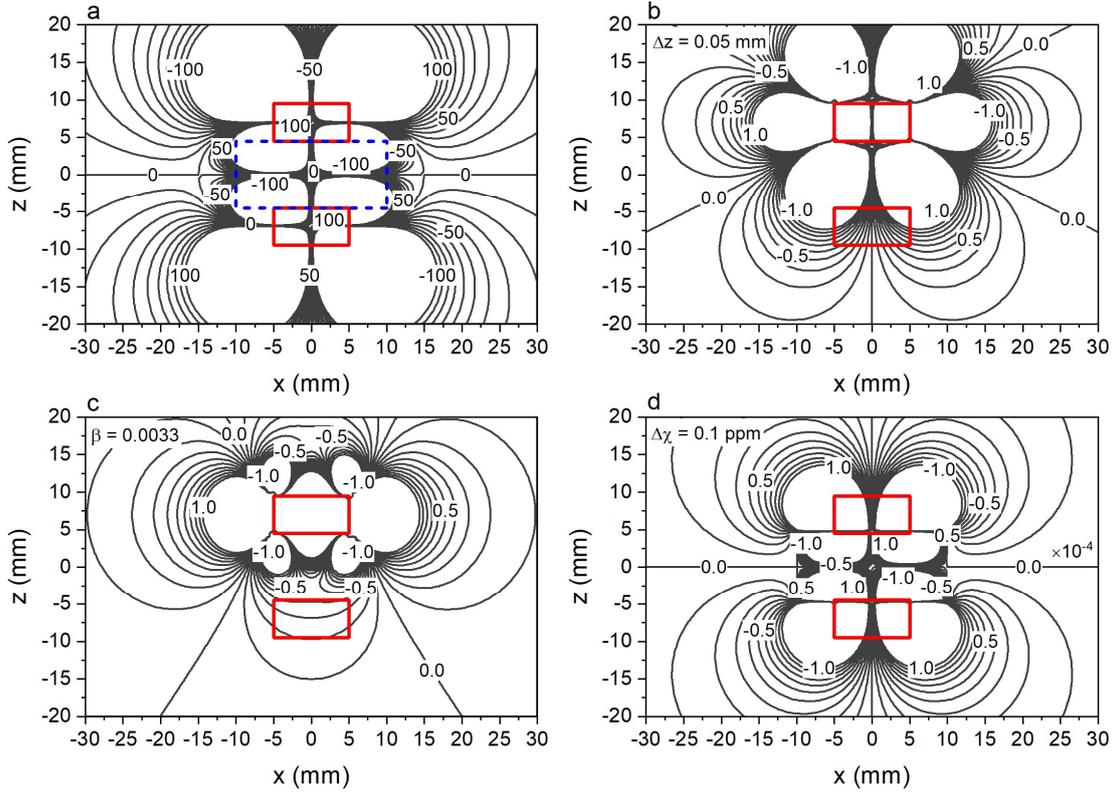

**FIG.5.** Contour plots of (a), $H_x$ in the range of ±100 Oe, (b), $\Delta H_x$ caused by vertical shift of the top magnet by 50 μm, (c), $\Delta H_x$ caused by a pitch rotation of 0.0033 rad, and (d), $\Delta H_x$ induced by change of blood susceptibility. Note that the unit of (a)-(c) is Oe, and that of (d) is $10^{-4}$ Oe.

**Numerical simulation of MDVS signal.** Bearing in mind these analytical results, we now calculate the contributions of each term in Eq. (1) numerically under typical measurement conditions using the Software for Multiphysics Simulation (COMSOL). To facilitate the discussion, we re-set the coordinate system as follows (as we now have two magnets): x-axis and y-axis are parallel and transverse to the finger, respectively, z-axis is perpendicular to the cylindrical magnet surface, and the coordinate origin is at the middle of the magnet with z-axis passing through the rotational center of both magnets. Figure 5a shows the calculated contour lines of $H_x$ on the xz plane in the range of ±100 Oe, superimposed with the cross-sections of the magnets and fingertip. The two small rectangular boxes denote the cylindrical magnets with thickness $T = 5$ mm, radius $R$ = 5 mm, and saturation magnetization $M_s$ = 962.9 kA/m. The large rectangular box defined by $|x| \leq 10$ mm and $|z| \leq 4.5$ mm represents the fingertip with $|y| \leq 5$ mm. The spacing between the bottom surface of the upper magnet and top surface of lower magnet is initially set at 9 mm. The bottom magnet is fixed, whereas the top magnet is movable in response to the skin movement. The magnetizations of both magnets are pointing in the positive z-direction.

To maximize the detection signal, the sensor must be placed at a location where the field strength is below the sensor's dynamics range (criterion 1) and at the same time the change caused by sensor displacement or rotation is largest (criterion 2). For the particular senor used in the experiment which has a dynamic range of ±100 Oe, the simulation results in Fig.5a show that criterion 1 is satisfied as long as the sensor is placed in the region where the contour lines are shown ($|H_x| < 100$ Oe). However, when the sensor is too far away from the magnet, it may not be able to detect the small change of field caused by the blood flow, which may originate from magnetic displacement/rotation or blood susceptibility change, or the combination of all these factors. We first calculate the field change ($\Delta H_x$) caused by magnet displacement or rotation under typical measurement conditions. Figure 5b shows the contour of $\Delta H_x$ caused by the movement of the top magnet in z-direction by 50 $\mu$m, which corresponds to the typical displacement of magnet measured by the vibrometer due to blood flow. Similarly, Fig.5c shows the change caused by rotating the magnetization of the top magnet by $3.3 \times 10^{-3}$ rad away from z axis around y-axis (*i.e.*, $\beta = 0.19^o$). The rotation angle is chosen such that the maximum deflection of the edge of the magnet is comparable to the vertical displacement used for obtaining the results in Fig.5b. From both figures, we can see that maximum change appears along the middle line of the top magnet in x-direction with its magnitude decreasing quickly away from the magnet.

Within the range of $15\text{ mm} < |x| < 20\text{ mm}$ and $5\text{ mm} < z < 10\text{ mm}$, it is possible to obtain a change of 0.5 – 1 Oe for $H_x$, which is detectable by the sensor.

We next simulate the field change caused by susceptibility change of the blood. Due to the difference in magnetic responses of $HbO_2$ and $Hb$, the magnetic susceptibility of oxygenated and de-oxygenated blood varies slightly,[7] and its difference relative to water $\Delta\chi$ can be approximated as follows[28]:

$$\Delta\chi = Hct(\Delta\chi_{do}(1 - HbO_2) + \Delta\chi_{oxy}), \tag{4}$$

where $Hct$ is the volume fraction of red blood cell, $\Delta\chi_{oxy}$ is the susceptibility difference between fully oxygenated blood and water, $\Delta\chi_{do}$ is the susceptibility difference between fully oxygenated and deoxygenated red blood cells, and $HbO_2$ ranges from 0 (fully deoxygenated) to 1 (fully oxygenated)[29,28,30,31] According to Jain et al., $\Delta\chi/Hct$ is about $-2 \times 10^{-8}$ at fully oxygenated state and $2.5 \times 10^{-7}$ at deoxygenated stated. Since the magnetic susceptibility of water is $7.18 \times 10^{-7}$ and $Hct$ is around 0.4-0.5, the magnetic susceptibility of blood is between $2.8 \times 10^{-7} - 4.8 \times 10^{-7}$, depending on its oxygenation state. The maximum change is less than 0.1ppm. To simulate its effect on field distribution, we calculate the field difference caused by the variation of magnetic susceptibility in the fingertip region by 0.1 ppm, and results are shown in Fig.5d. As can be seen, the change in $H_x$ is on the order of $10^{-5}$ Oe in the region of interest, *i.e.*, $15\text{ mm} < |x| < 20\text{ mm}$ and $5\text{ mm} < z < 10\text{ mm}$, which is too small to be detected by the magnetic sensor used in this work (note: the unit in Fig.5d is $10^{-4}$ Oe). The above results demonstrate clearly that the pulsatile signals detected by the magnetic sensor are dominantly of mechanical origin, the MMSB contribution, if any, is negligible.

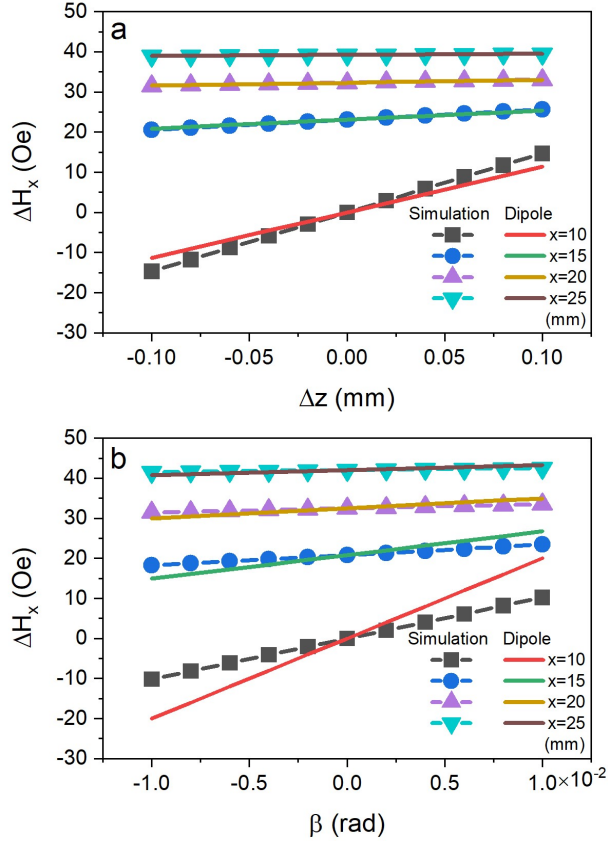

**FIG.6.** Comparison of field change obtained by COMSOL simulation (symbol) and dipole approximation (solid-line): (a), $\Delta H_x$ as a function of vertical shift $\Delta z$ and (b), $\Delta H_x$ as a function of pitch angle $\beta$ of the top magnet at different radial position, $x = 10, 15, 20, 25$ mm, from the center of the magnet.

Before we end this section, we discuss the validity of the dipole approximation by comparing the analytical and COMSOL simulation results. Figure 6a shows $\Delta H_x$ obtained by the two methods as a function of vertical displacement of the top magnet ($\Delta z$) at different radial position ($x$) along the middle line of the magnet. Similar results are shown in Fig.6b for pitch angle $\beta$. Solid-lines are results obtained from the dipole approximation and symbols are from COMSOL simulations. As expected, good agreements are obtained when both $\Delta z$ and $\beta$ are small, and $x$ is about 3 times larger than that of the radius of the magnet. Therefore, the dipole approximation can be used to optimize the position of the sensor with respect to the magnet.

**Discussion**

We have proposed and designed a pulse detection device which allows us to conduct a comparative study to determine the origin of so-called MPG signals through simultaneous measurements using a magnetic sensor and a vibrometer. Based on both experimental and simulation results, we conclude unambiguously that the detected signal is not originated from change of blood magnetic properties but rather from mechanical motion of the magnet induced by blood flow, including both vertical displacement and rotation. Apart from the determination of signal origin, the proposed fingertip-type magnetically detected vibration sensor is much more robust and simpler than the previously report wrist-type MPG sensor, and thus is more suitable for practical applications. It has clear advantages over existing pulse detectors in coping with unusual monitoring conditions. In addition to heartrate detection, its direct relevance with the blood pressure wave makes it promising for blood pressure measurement.

**Methods**

**Signal processing and statistical analysis.** The detected signals are processed with MATLAB as follows. First, the raw data with a sampling rate of 240Hz is segmented into sequences with a duration of 30s. Second, a bandpass filter with a passband of 0.8Hz - 10Hz is used to remove high-frequency noise and baseline shift. Third, z-score analysis is used to transform two types of signals with different amplitudes and units into similar range, as shown in Figure 2a, 2d and 2g. Fourth, single pulse is extracted by averaging all pulses of the period of data sequence, and the results are shown in Figure 2b, 2e and 2h. Figure 2c and 2f are second derivative of the single pulse signals. Finally, Bland-Altman plot is analyzed based on 60 pulses from 3 sequences of signals.

**Simulation setup.** Simulation of magnetic signals was performed using COMSOL Multiphysics® software. The initial distance between the body centers of the two magnets is set at 14 mm with a finger diameter of 9 mm. When simulating the effect of vertical vibration, only the position of magnet in positive z-axis is varied to simulate the expansion of finger due to blood circulation. A cuboid is used to simulate the susceptibility change inside the finger. When simulating the effect of rotation, only magnet in positive z-axis is rotated around the center of bottom surface.

**Acknowledgements**

This work is supported by the Advanced Research and Technology Innovation Centre (ARTIC), the National University of Singapore under Grant A-0005947-23-00 and NUS GAP Funding (RIE2025) under Grant A-8000656-00-00. The authors thank Dr. Qi Zhang for his help in the initial stage of the study.

**Author Contributions**

Y.H.W. conceived the idea and supervised the project. S.H.X. designed and fabricated the prototype and performed the COMSOL simulation. S.H.X. and Y.H.W. performed the measurements and analysed the data. G. X. Y. helped with the measurements. S.H.X. and Y.H.W. wrote the manuscript.

**Competing interests**

The authors declare no competing interests.

**Data Availability**

The data that support the findings of this study are available from the corresponding author upon reasonable request.